\documentclass{ws-procs9x6}
\newcommand{\ba}{\begin{eqnarray}}
\newcommand{\ea}{\end{eqnarray}}
\newcommand{\bmath}{\begin{mathletters}}
\newcommand{\emath}{\end{mathletters}}
\newcommand{\ban}{\begin{eqnarray*}}
\newcommand{\ean}{\end{eqnarray*}}

\newcommand{\bsub}{\begin{subequations}}
\newcommand{\esub}{\end{subequations}}

\def\ket#1{|#1\rangle}
\def\bra#1{\langle#1|}

\def\bz{\beta_0}

\begin{document}

\title{Coexistence of order and chaos at critical points of first-order 
quantum phase transitions in nuclei}
\author{M. Macek and A. Leviatan}
\address{Racah Institute of Physics, The Hebrew University, 
Jerusalem 91904, Israel}

\begin{abstract}
We study the interplay between ordered and chaotic dynamics at the critical 
point of a generic first-order quantum phase transition 
in the interacting boson model of nuclei. Classical and quantum analyses 
reveal a distinct behaviour of the coexisting phases. While the dynamics 
in the deformed phase is robustly regular, the spherical phase 
shows strongly chaotic behavior in the same energy intervals. 
The effect of collective rotations on the dynamics is investigated. 
\end{abstract}

\bodymatter
\vspace{1 em}
Quantum phase transitions (QPTs) are structural changes occurring 
at zero temperature, resulting from a variation of parameters 
in the quantum Hamiltonian. They have become a topic of great interest 
in diverse many-body systems, {\it e.g.}, atoms, molecules and nuclei. 
The abrupt changes in the the system's ground state
affect the nature of the underlying dynamics and can lead 
to the emergence of quantum chaos. 
In the present contribution, we examine this effect 
in generic (high-barrier) first-order QPTs in nuclei, 
focusing on the role of the barrier separating the two coexisting 
phases~\cite{ref:MacLev11}. 

We employ the interacting boson model (IBM)~\cite{ref:Iach87}, 
widely used in the description of quadrupole 
collective states in 
nuclei, in terms of a system of $N$ monopole ($s$) and
quadrupole ($d$) bosons, representing valence nucleon pairs.
A geometric visualization of the model 
is obtained by an energy surface, $V(\beta,\gamma)$, which serves as a 
Landau potential. The latter is defined by the expectation value of the 
Hamiltonian in the intrinsic condensate state 
$\vert\beta,\gamma ; N \rangle = 
(N!)^{-1/2}[\Gamma^{\dagger}_{c}(\beta,\gamma)]^N\vert 0\rangle$,  
where $\Gamma^{\dagger}_{c}(\beta,\gamma) = 
{\textstyle\frac{1}{\sqrt{2}}}
[\beta\cos\gamma d^{\dagger}_{0} + \beta\sin{\gamma} 
{\textstyle\frac{1}{\sqrt{2}}}( d^{\dagger}_{2} + d^{\dagger}_{-2}) 
+ \sqrt{2-\beta^2}s^{\dagger}]$. 
The quadrupole shape parameters in the 
intrinsic state characterize the associated equilibrium shape. 
QPTs have been extensively studied in the 
IBM~\cite{ref:Iac11,ref:Cejn10}. 
To construct a critical Hamiltonian it is convenient to resolve it 
into intrinsic and collective parts~\cite{ref:Lev87},
\ba
\hat{H}_{\mathrm{cri}} = \hat{H}_{\mathrm{int}} 
+ \hat{H}_{\mathrm{col}} ~.
\label{Hcri}
\ea
The intrinsic part ($\hat{H}_{\mathrm{int}}$) determines the potential 
$V(\beta,\gamma)$, while the collective part ($\hat{H}_{\mathrm{col}}$)
contains kinetic rotational terms which do not affect its shape. 
For a first-order critical point, the two parts of the full critical 
Hamiltonian ($\hat{H}_{\mathrm{cri}}$) can be transcribed in the 
form~\cite{ref:Lev06} 
\bsub
\ba
\hat{H}_\mathrm{int} &=& 
\bar{h}_2 P^\dag_2(\beta_0)\cdot\tilde{P}_2(\beta_0) ~,
\label{Hint}
\\
\hat{H}_\mathrm{col} &=& 
\bar{c}_3 \left [\, \hat{C}_{O(3)} - 6\hat{n}_d \,\right ]
+ \bar{c}_5 \left [\, \hat{C}_{O(5)} - 4\hat{n}_d \,\right ]
+ \bar{c}_6 \left [\, \hat{C}_{\overline{O(6)}} - 5\hat{N}\,\right ] ~. 
\qquad
\label{Hcol}
\ea
\esub
Here $P^{\dagger}_{2\mu}(\beta_0) = 
\beta_{0}\,s^{\dagger}d^{\dagger}_{\mu} + 
\sqrt{7/2}\,\left( d^{\dagger} d^{\dagger}\right )^{(2)}_{\mu}$, 
$\hat{n}_d$ ($\hat{N}$) is the $d$-boson (total) number operator and 
$\hat{C}_{G}$ denotes the quadratic Casimir of the group $G$ as 
defined in Ref.~\refcite{ref:Lev87}. 
Barred parameters imply scaling by $N(N-1)$.
For $\bz>0$, $\hat{H}_\mathrm{cri}$ annihilates 
both the spherical $s$-condensate, 
$\vert\beta=0,\gamma ; N \rangle$, and the deformed condensate, 
$\vert\beta=\beta_{\mathrm{e}}=\sqrt{2}\bz(1+\bz^2)^{-1/2},
\gamma=0 ; N\rangle$ 
which correspond to the two coexisting shape phases of the 
nucleus. 

The classical limit of the IBM Hamiltonian 
is obtained through the use of Glauber coherent states 
and taking  $N\rightarrow\infty$~\cite{ref:AW93}. 
The resulting classical image 
of $\hat{H}_\mathrm{cri}$ contains complicated expressions, including 
square roots of polynomials in the coordinates $\beta,\gamma$ and 
their conjugate momenta $p_{\beta},p_{\gamma}$. 
Setting all momenta to zero, leads to the potential
\ba
\label{eq:Vcl}
V_\mathrm{cri}/h_2 &=& 
{\textstyle\frac{1}{2}}\bz^2 \beta^2  
+ \textstyle{\frac{1}{4}}(1-\bz^2)\beta^4 
- \textstyle{\frac{1}{2}}\bz \sqrt{2-\beta^2} \beta^3\cos{3\gamma} ~.
\ea
The potential has a spherical minimum at $\beta=0$ 
degenerate 
with a prolate-deformed 
minimum at ($\beta=\beta_{\mathrm{e}},\gamma=0$) 
both at energy $V_\mathrm{min}=0$. 
The barrier, separating the two minima, is located at   
$\beta=[1-(1+\bz^2)^{-1/2}]^{1/2}$ and has a height 
$V_b=h_{2}[1-(1+\beta_0^2)^{1/2}]^2/4$. 
The limiting value at 
the domain boundary is $V_{\mathrm{cri}}(\beta=\sqrt{2},\gamma)=h_2$. 
The potential can also be expressed in Cartesian coordinates 
$x=\beta\cos{\gamma}$ and $y=\beta\sin{\gamma}$.
\begin{figure}[t]
\begin{center}
\epsfig{file=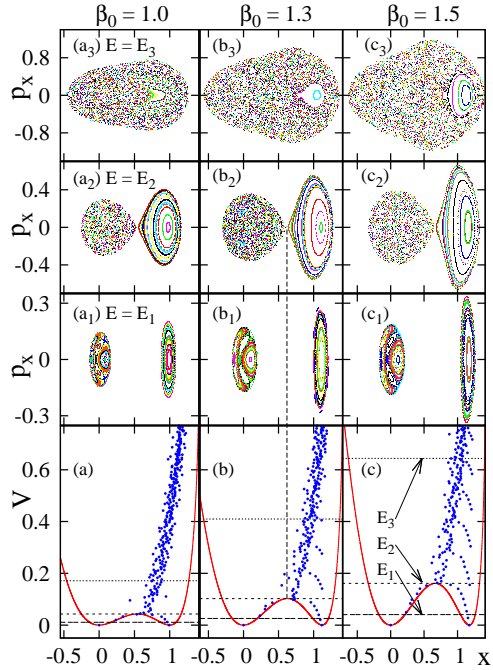,width=0.56\linewidth}
\end{center}
\caption{Poincar\'e sections at energies 
$E_k$ (panels $a_k$-$b_k$-$c_k$, $k=$ 1-3)
and Peres lattices of ($N=80,L=0$) eigenstates of 
$\hat{H}_\mathrm{int}~(\ref{Hint})$ overlayed on the classical potential 
$V_\mathrm{cri}(x,y=0)$  for $\beta_0=1.0,1.3,1.5$ and $h_2=1$. 
Notice the well-separated regular and chaotic dynamics associated with 
the deformed ($x>0$) and spherical ($x=0$) minima, respectively.}
\label{fig:1}
\end{figure}
\begin{figure}[t]
\begin{center}
\epsfig{file=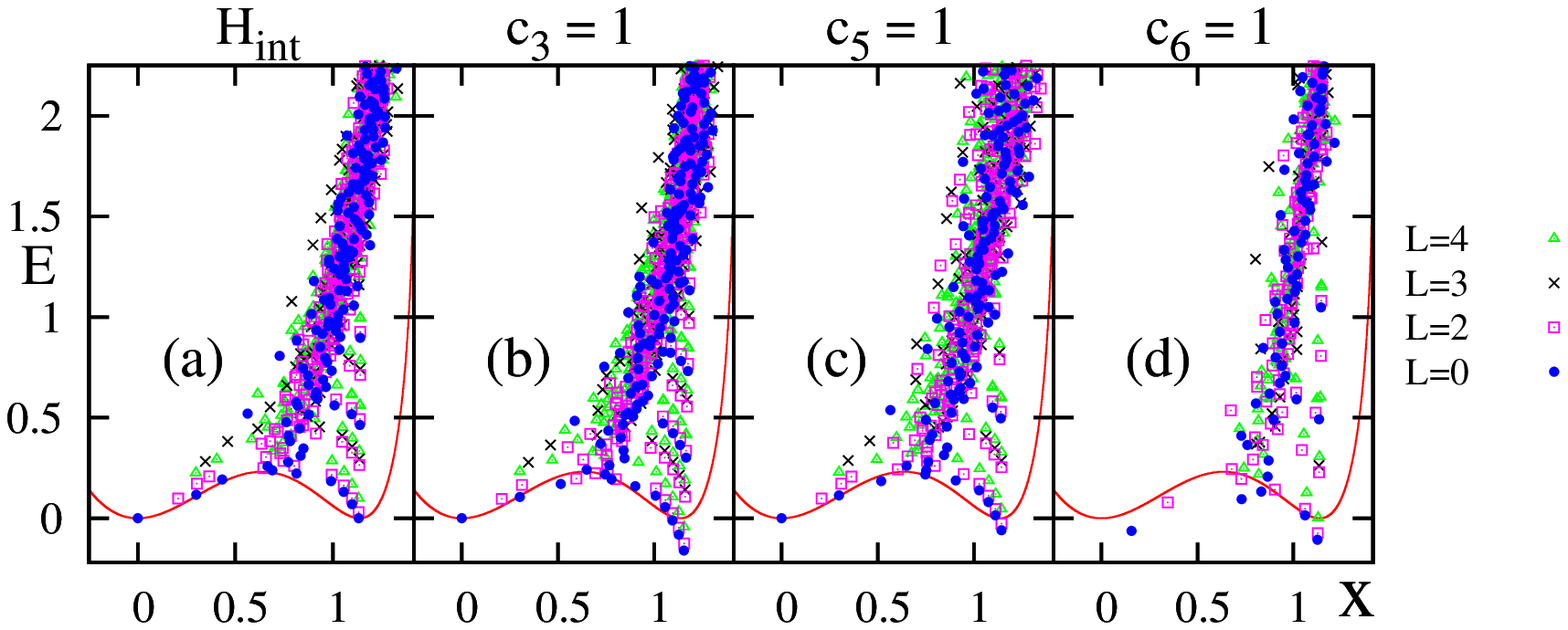,width=0.75\linewidth}
\end{center}
\caption{Peres lattices for eigenstates 
with $N=50,\,\beta_0=1.35,\,h_2=1$ and $L=0,2,3,4$ of 
$\hat{H}_\mathrm{int}~(\ref{Hint})$ (panel $a$) and additional 
collective terms~(\ref{Hcol}) involving 
$O(3),\,O(5)$ and $\overline{\mathrm{O(6)}}$ rotations 
(panels $b$-$c$-$d$), respectively. 
Notice the well-defined rotational bands 
($K=0$, $L=0,2,4$) and ($K=2$, $L=2,3,4$)
formed by the regular states in the deformed phase
shown in panels $(a,b,c)$, which are distorted in panel $(d)$.}
\label{fig:2}
\end{figure}

Chaotic properties of the IBM have been 
extensively studied~\cite{ref:AW93}, albeit, with a simplified Hamiltonian, 
giving rise to an extremely low, hence non-generic, barrier. 
To study the effect of the barrier, 
we consider the classical 
dynamics associated with $\hat{H}_\mathrm{int}$~(\ref{Hint}) 
constrained to $L=0$. 
The Poincar\'e sections 
for $\bz=1.0,1.3,1.5$, 
are displayed in Fig.~\ref{fig:1}. The three cases correspond to 
``low'', ``medium'' and ``high'' potential barriers 
$V_b/h_2=$ 0.04, 0.10, 0.16, 
(compared to $V_b/h_2=$ 0.0009 in previous works~\cite{ref:AW93}).
The sections are plotted at energies $E_1=V_b/4$, $E_2=V_b$ and $E_3=4 V_b$. 
In all cases, the motion is predominantly regular 
at low energies and gradually turning chaotic as the energy increases. 
However, the classical dynamics evolves 
differently in the vicinity of the two wells.
The family of regular trajectories 
near the deformed minimum 
forms a simple pattern of concentric loops around 
a single stable (elliptic) fixed point. The trajectories remain regular even 
at energies $E \gg V_b$ (panels $a_3$-$b_3$-$c_3$). 
In contrast, near the spherical minimum, a more complex 
regular dynamics at low energies (panels $a_1$-$b_1$-$c_1$),
turns chaotic at a much lower energy ($E\approx V_b/3$) and complete 
chaoticity is reached near the barrier top.  
The clear separation between regular and chaotic dynamics, 
associated with the two minima, persists all the way to $E=V_b$, 
(panels $a_2$-$b_2$-$c_2$). In general, 
the regularity is more pronounced for higher barriers (larger $\bz$).
These attributes are present also in the quantum analysis in terms of 
Peres lattices~\cite{ref:Pere84}, formed by the set of 
points $\{x_i,E_i\}$, with $x_i = \sqrt{2 \bra{i}\hat{n}_d\ket{i}/N}$, 
and $E_i$ the energy of the eigenstate $\ket{i}$. 
The Peres lattices for $L=0$ eigenstates of 
$\hat{H}_\mathrm{int}$~(\ref{Hint}) 
are shown in the bottom row in panels ($a$-$b$-$c$), nested within 
the Landau potential $V(x,y=0)$. 
For each $\beta_0$, one can clearly identify several regular sequences of 
states localized in and above the respective deformed wells. 
A close inspection reveals that their $x_i$-values lie in the intervals 
of $x$-values occupied by the regular tori in the Poincar\'e sections. 
In contrast, the remaining states, including those residing in the 
spherical minimum, do not show any obvious patterns and lead to 
disordered (chaotic) meshes of points at high energy $E>V_b$. 

Fig.~\ref{fig:2} displays combined Peres lattices 
for eigenstates with $N=50$, $\beta_0=1.35$ and $L=0,2,3,4$ of 
$\hat{H}_\mathrm{int}~(\ref{Hint})$ (panel $a$) 
supplemented with three different additional collective terms~(\ref{Hcol})
$\{c_3,c_5,c_6\}=\{1,0,0\},\{0,1,0\}$ and $\{0,0,1\}$ in panels 
($b$,$c$,$d$), respectively. 
The regular sequences of $L=0$ eigenstates, mentioned previously in 
relation to Fig.~\ref{fig:1}, are seen to be bandhead states of 
($K=0$, $L=0,2,4$) rotational bands of states with nearly equal values 
of $\langle \hat{n}_d \rangle$.
Similarly, sequences of $L=2,3,4$ states form $K=2$ bands. 
Such ordered band structures are not present in the chaotic parts of 
the lattice. The ordered band structure is well preserved by the 
$O(3)$ and $O(5)$ terms but is suppressed by the 
$\overline{\mathrm{O(6)}}$ term.

This work is supported by the Israel Science Foundation.

\end{document}